\documentclass[%
 aps,
 prl,%
 amsmath,amssymb,
preprint,%
superscriptaddress,showpacs]{revtex4-1}

\usepackage{color, graphicx}
\usepackage{dcolumn}
\usepackage{bm}
\usepackage{amssymb}
\usepackage{latexsym}
\usepackage{amsfonts}
\usepackage{amsmath}
\usepackage{multirow}
\usepackage{ifthen}

\begin{document}


\title{Control of acoustic absorption in 1D scattering by indirect coupled resonant scatterers}

\author{A. Merkel}
\author{G. Theocharis}
\author{O. Richoux}
\author{V. Romero-Garc\'ia}
\author{V. Pagneux}
\affiliation{LUNAM Universit\'e, Universit\'e du Maine, LAUM UMR CNRS 6613, Av. O. Messiaen, 72085 Le Mans, France}


\begin{abstract}
We experimentally report perfect acoustic absorption through  the interplay of the inherent losses and transparent modes with high $Q$ factor. These modes are generated in a two-port, one-dimensional waveguide which is side-loaded by isolated resonators of moderate $Q$ factor. In symmetric structures, we show that in the presence of small inherent losses, these modes lead to coherent perfect absorption associated with one-sided absorption slightly larger than 0.5. In asymmetric structures, near perfect one-sided absorption is possible (96 \%) with a deep sub-wavelength sample ($\lambda/28$). The control of strong absorption by the proper tuning of few resonators with weak losses will open new possibilities in various wave-control devices.
\end{abstract}

\pacs{43.20.Mv, 43.20.Hq,43.20.Fn}

\maketitle



One of the most inspiring outcomes in the field of atom optics is the electromagnetically-induced-transparency (EIT) which results from the coherent interferences between different excitation pathways of the excited states \cite{boller1991}. 
In the recent years, and after the theoretical revealing of the similarities between atoms driven by optical fields and resonators excited by incident waves, there has been an increasing interest in the implementations of classical analogues of EIT-like spectra. Thus, EIT-like behaviors have been theoretically studied and experimentally observed in plasmonic, photonic and acoustic resonator systems (see \cite{boudouti2008,Miro2010,Boris2010,santillan2011,tan2012,santillan2014,Peng2014,mouadili2014} and references therein). 
In the lossless case, these open system configurations show transparent peaks, with strongly dispersive and enhanced slow wave propagation, originated from the coherent interference between two or more coupled resonators. The $Q$ factor of these transparent resonances is much higher than the $Q$ factor of each individual resonator. 
In the experiments, the inherent losses result to a degradation of both the wave transparency and the slowdown propagation (see \cite{Baba2008} for optics and \cite{theocharis2014} for acoustics). Thus, in many fields of wave physics, the presence of losses has been dealt as a negative feature and for instance in photonics, many efforts are focused on how to minimize these. 

However, in some cases, losses could lead to interesting and useful wave phenomena. Among these are the critical coupling mechanism \cite{Slater1950, Xu2000, Cai2000} and Perfect Absorption (PA). The former comes from the balance between the leakage rate of energy out of the resonator (resonator-waveguide coupling) and the resonator losses. This balance leads to a maximum of energy absorption at the resonance frequency \cite{Xu2000}. 
In optics, different configurations including Bragg reflectors \cite{Tischler2006},  graphene-based hyperbolic metamaterials \cite{Xiang2014} or photonic crystal slabs \cite{piper2014} have been used to obtain nearly PA of the incident light. In acoustics, PA has been only recently obtained in an one-port system by the use of sub-wavelength resonators in front of a rigid backing \cite{GuancongMa2014,Romero2015,Leroy2015}. 
The importance of the PA using acoustic resonators concerns the design of sub-wavelength sound absorbers. 
PA has been also numerically observed in a two-port system by controlling the coherence of the two-sided incident waves on the target material \cite{Wei2014APL}. This is an application of the Coherent Perfect Absorption (CPA) mechanism \cite{chong2010PRL} which results from an interplay of wave interference and absorption. However, PA in a two-port system with one-sided incident wave is less explored.  

In this letter, we experimentally show that CPA can be achieved by tuning the leakage of an EIT-like mode in the presence of small inherent losses. This mode is generated by two indirect coupled Helmholtz Resonators (HR), each of them being alone an inefficient absorber. 
The EIT-like mode comes from the interaction of two or more complex resonance poles of the scattering coefficients in the complex frequency plane. For one-sided incident wave, we show that,  only due to the viscothermal effects at the walls, this mode can be critically coupled leading to a maximum of absorption. The critical coupling is achieved by simply modifying the geometry of the sample. In asymmetric structures, near perfect one-sided absorption is possible. In the following, we will consider a pair of HRs in three different configurations: point symmetric, mirror symmetric and asymmetric scatterers. 

\begin{figure}[ht!]
\includegraphics[width=8.6cm]{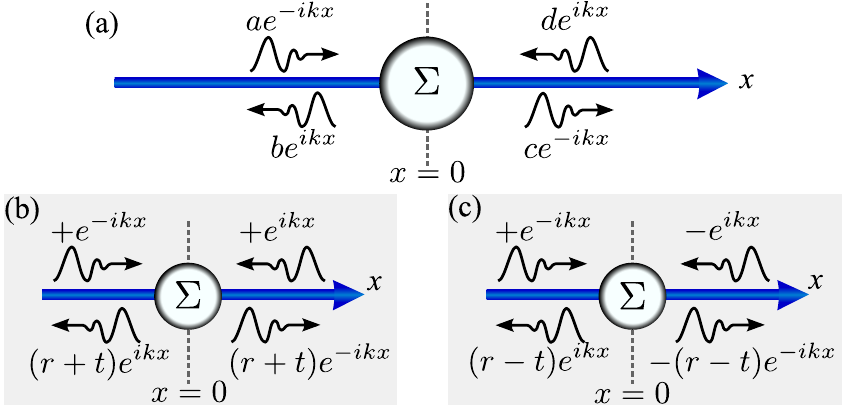}
\caption{(Color online)  (a) Schematic of the two-port scattering process. (b) Symmetric and (c) antisymmetric uncoupled sub-problems for the case of a mirror-symmetric scatterer $\Sigma$. The time convention is $e^{i\omega t}$. }
 \label{fig:theory} 
\end{figure}

Let us consider a two-port, one-dimensional and reciprocal scattering process. The relation between the amplitudes of the incoming ($a$, $d$), and outcoming ($b$, $c$) waves, on both sides of the scatterer $\Sigma$ in Fig. \ref{fig:theory}(a), is given by
\begin{eqnarray}
 \left( \begin{array}{c} c \\ b \end{array} \right) & = & \textbf{S}(f) \left( \begin{array}{c} a \\ d \end{array} \right) 
   =   \left( \begin{array}{cc}
  t & r^{R} \\
  r^{L} & t
  \end{array} \right)\left( \begin{array}{c} a \\ d \end{array} \right) ,
    \label{smatrix} 
\end{eqnarray}
where $\textbf{S}(f)$ is the scattering matrix (S-matrix), $f$ is the incident wave frequency, $t$ is the complex transmission coefficient, $r^{L}$ and $r^{R}$ are the complex reflection coefficients for left $L$ and right $R$ incidence, respectively. The eigenvalues of the S-matrix are expressed as $\xi_{1,2}=t\pm [r^{L}r^{R}]^{1/2}$ and the ratio of the eigenvector components $v_1$ and $v_2$ is $v_2/v_1=\pm(r^L/r^R)^{1/2}$. A zero eigenvalue of the S-matrix corresponds to the case in which the incident waves can be completely absorbed ($b=c=0$). This, called CPA  \cite{chong2010PRL}, happens when $t=\pm [r^{L}r^{R}]^{1/2}$ and the incident waves $a,d$ correspond to the relevant eigenvector.\\
{\it Mirror symmetric scatterer}.--- If the scatterer $\Sigma$ is mirror symmetric, $r^{R}=r^{L}\equiv r$ and the problem can be reduced to two uncoupled sub-problems by choosing incident waves that are symmetric [see Fig. \ref{fig:theory}(b)] or antisymmetric [see Fig. \ref{fig:theory}(c)] with the reflection coefficients $r_s=r+t$ and $r_a=r-t$. 
In particular, the reflection and transmission coefficients of the initial problem in Fig. \ref{fig:theory}(a) can be expressed as $r= (r_s + r_a )/2$, and $t=(r_s - r_a )/2$ while the eigenvalues of the S-matrix can be written as  $\xi_1=r_s$ and $\xi_2=-r_a$. 
For an one-sided incident wave, the absorption coefficient defined as 
$\alpha= 1-  |r|^2 - |t|^2 $ becomes $\alpha= (\alpha_s + \alpha_a )/2$, where $\alpha_s \equiv 1-  |r_s|^2$ and $\alpha_a \equiv 1-  |r_a|^2$. 
Achieving $\alpha(f_{max})=1$ at a frequency $f_{max}$, is equivalent to getting simultaneously the minima of the reflection coefficients of the two sub-problems, i.e., $r_a(f_{max})=r_s(f_{max})=0$ [$\alpha_s(f_{max})=\alpha_a(f_{max})=1$]. This has been achieved in \cite{piper2014} for a mirror symmetric slab through intensive numerical calculations. \\
{\it Point symmetric scatterer}.--- We now further consider that $\Sigma$ is a point scatterer, i.e., its length range is reduced to $x=0$. This case is also relevant to the study of absorption by deep sub-wavelength scatterers. Imposing the continuity of the wave-field at this point \cite{supplement} yields $1+r=t$. This corresponds to $r_a=-1$, i.e., the scatterer $\Sigma$ is transparent to antisymmetric incident waves, $\alpha_a=0$,  and thus $\alpha= \alpha_s/2 \leq 1/2$. The maximum value of one-sided absorption, appeared at $f_{max}$, is $\alpha(f_{max})=1/2$. It corresponds to $\alpha_s(f_{max})=1$ which gives  $r_s(f_{max})=\xi_1(f_{max})=0$, and $r(f_{max})=-t(f_{max})=1/2$. Thus,  $\alpha(f_{max})=1/2$ corresponds to CPA to the two-incident waves problem for symmetric and in phase incoming waves at $f_{max}$, i.e., $v_2(f_{max})/v_1(f_{max})=1$ \cite{Wei2014APL}. 
  
The experimental setup \cite{supplement} which is used to study the control of absorption by point symmetric, mirror symmetric and asymmetric scatterers, consists of a cylindrical air-filled waveguide of radius $R_w=2.5$~cm with two HRs. The HRs, with a moderate quality factor $Q^{HR}$ ($50\sim90$) \cite{supplement}, are composed of a cylindrical neck of radius $R_n=1$~cm and length $L_n=2$~cm, and a cylindrical volume of  radius $R_v=2.15$~cm and variable length $L_v$. The resonance frequency of the two HRs $f^{HR}_{1,2}$ can be tuned with the variable length $L_v$. The absorption obtained by one HR alone is $\alpha_{HR}\simeq0.1$. 
The incident wavelengths ($\lambda>30$~cm) are much larger than $R_n$ and $R_w$, so that the resonators are assumed to be point scatterers and the problem is considered as one-dimensional. 

\begin{figure}[ht!]
\includegraphics[width=8.6cm]{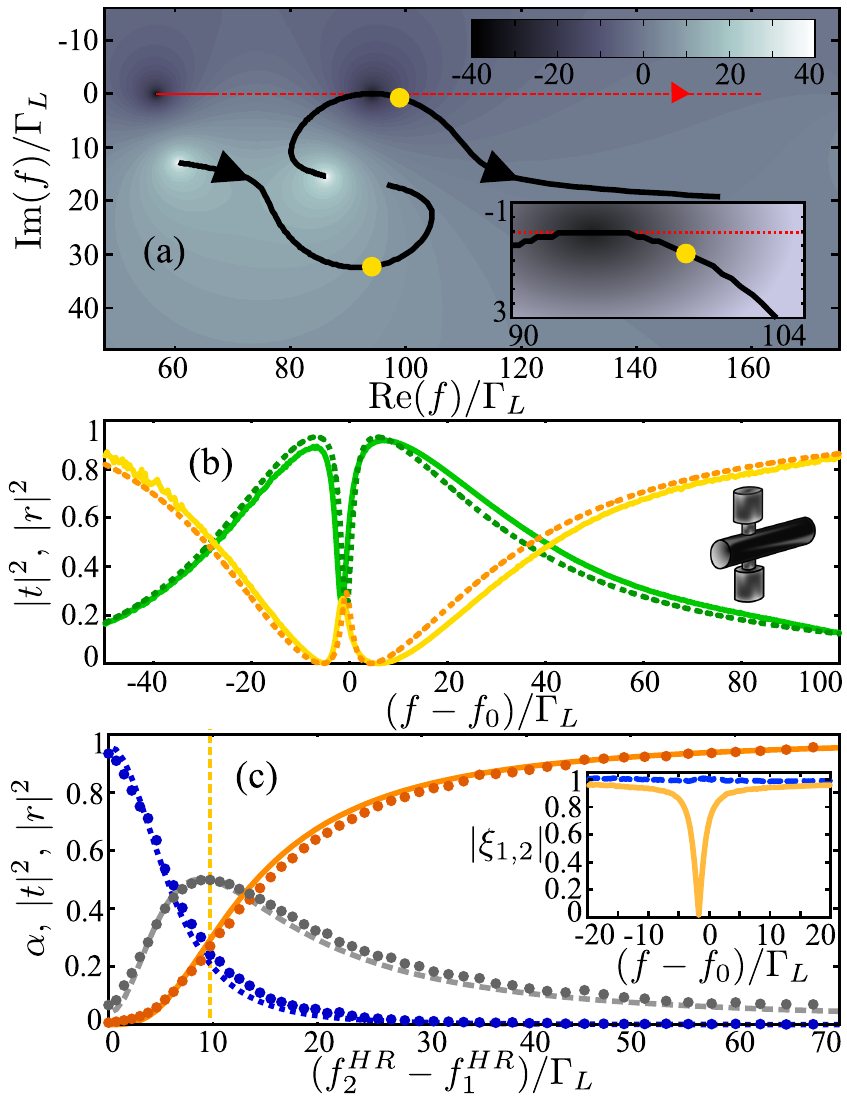}
\caption{(Color online) Point symmetric scatterer. 
(a) Theoretical lossless transmission coefficient (logscale) in the complex frequency plane with the trajectories of the poles (black lines) and of the zero (red dashed line) for $(f^{HR}_2-f^{HR}_1)/\Gamma_L \in [-37;67]$. The positions of the poles are marked for $(f^{HR}_2-f^{HR}_1)/\Gamma_L=9.8$ (yellow dots) and the underlying colormap image corresponds to $(f^{HR}_2-f^{HR}_1)/\Gamma_L=-37$. Inset: zoom around the critical coupling point. 
(b) Theoretical with losses (dashed curves), experimental (continuous curves) transmission (yellow curves) and reflection (green curves) coefficients vs frequency for $(f^{HR}_2-f^{HR}_1)/\Gamma_L=9.8$ with $f_0=311$~Hz. 
(c) Absorption (gray), transmission (orange) and reflection (blue) coefficients at $f_{max}$ vs $(f^{HR}_2-f^{HR}_1)/\Gamma_L$ (curves for theory and dots for experiments). 
Inset:  Eigenvalues of the experimental S-matrix for $(f^{HR}_2-f^{HR}_1)/\Gamma_L=9.8$. }
 \label{fig:parallel} 
\end{figure}

In our setup, a point symmetric scatterer consists of two detuned HRs ($f^{HR}_{1}\neq f^{HR}_{2}$) located at the same axial position [see the sketch in Fig. \ref{fig:parallel}(b)]. We define the detuning parameter as $(f_2^{HR}-f_1^{HR})/\Gamma_L$, where $\Gamma_L=3.14$~Hz is the decay rate due to losses of the HRs \cite{supplement}. 
Ignoring the losses, it has been shown that, for small values of the detuning parameter, the transmission coefficient presents an EIT-like mode with unity transmission at $f_0=(f^{HR}_2+f^{HR}_1)/2$ \cite{Xu2006,Yang2009,boudouti2008, santillan2011, mouadili2014}. 
The lossless transmission coefficient in the complex frequency plane is displayed in Fig. \ref{fig:parallel}(a);
this reveals two poles which are hybridized resonances resulting from the two resonances of the HRs.
As the detuning parameter changes, in our case by increasing the resonance frequency of one HR while keeping the other one fixed at $f^{HR}_1=294$~Hz,  the poles move [sense of the arrows in Fig. \ref{fig:parallel}(a)], interact and repel each others.
As $f_2^{HR}-f_1^{HR}\rightarrow 0$, one pole at $f_{pole}$, with $\text{Re}(f_{pole})\simeq f_0$, approaches the real axis giving rise to an EIT-like mode with high $Q$ factor. 
In other words, the interaction of the two resonances leads to a dark mode (the EIT-like mode) and a bright mode (with a corresponding pole far from the real axis). 
We now analyze the experimental results and compare them with the theoretical predictions taking into account the losses \cite{supplement} by looking at the scattering coefficients as a function of the frequency (corresponding to the real axis in the complex frequency plane). For small detuning parameter, the viscothermal losses importantly reduce the amplitude of the transparent peak associated with the EIT-like mode: instead of the unity transmission in the lossless case, the peak can take values between $0$ and $1$. 
For each value of the detuning parameter, the transparent peak is associated with a peak of absorption found at $f_{max}$ (slightly different from $f_0$ due to the losses). 
The maximum peak of absorption is found with $(f_2^{HR}-f_1^{HR})/\Gamma_L=9.8$, highlighted with circles in Fig. \ref{fig:parallel}(a), at $f_{max}=306~Hz$.
The relevant scattering coefficients are displayed in Fig. \ref{fig:parallel}(b). According to the theoretical considerations on point symmetric scatterers, $|t(f_{max})|=|r(f_{max})|=0.5$ corresponds to $\alpha_s(f_{max})=1$ and consequently to $\alpha(f_{max})=0.5$. This one-sided incident wave maximum of absorption is found when the leakage of the  EIT-like mode is tuned in order to balance the inherent losses, i.e., it is critically coupled \cite{Xu2000,bliokh2008,supplement,Leroy2015}. 
This is confirmed in Fig. \ref{fig:parallel}(c) where the scattering coefficients and the absorption at $ f_{max}$ are plotted as a function of the detuning parameter. In addition, one experimental eigenvalue becomes zero at $f_{max}$ as shown in Fig. \ref{fig:parallel}(c). As mentioned above, this is the symmetrical CPA \cite{Wei2014APL}; the case at which the incident waves from the two sides of the sample, corresponding to the S-matrix eigenvector such as $v_2/v_1=1$, are completely absorbed. 


\begin{figure}[ht!]
\includegraphics[width=8.6cm]{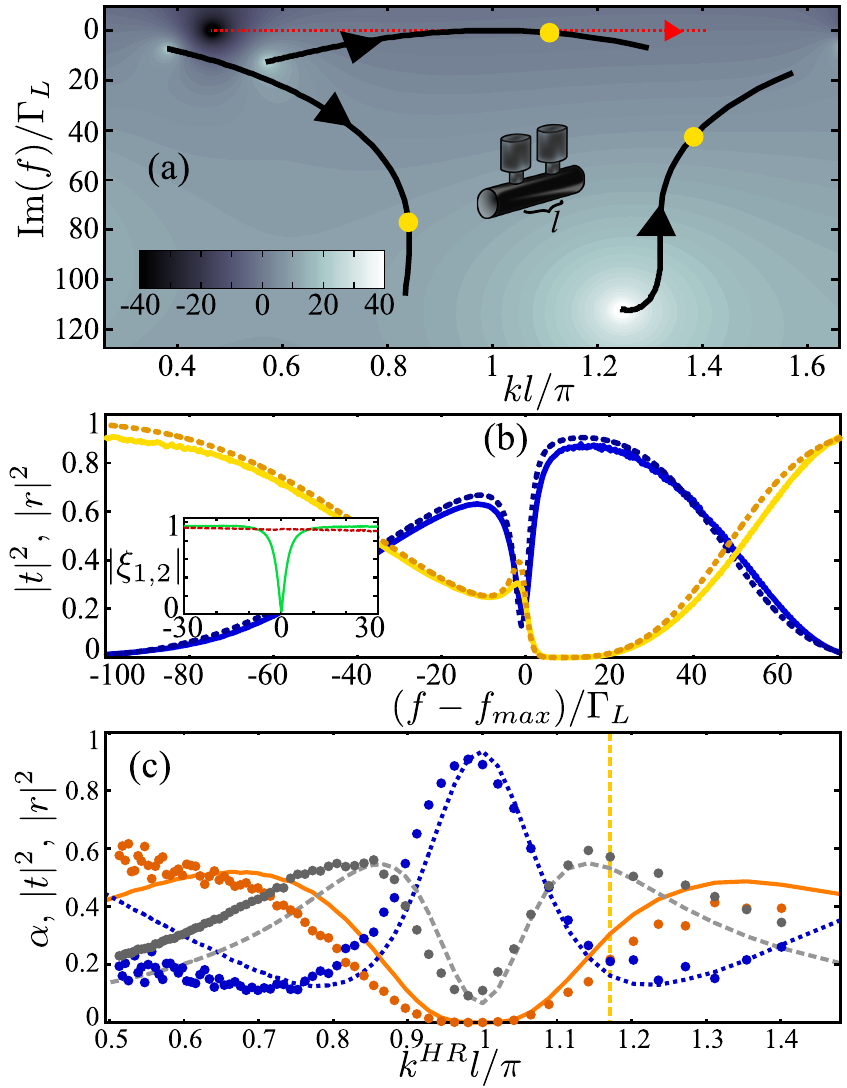}
\caption{(Color online) Mirror symmetric scatterer, $l=30$~cm. 
(a) Theoretical lossless transmission coefficient (logscale) in the complex frequency plane with the trajectories of the poles (black lines) and of the zeros (red dashed line) with $k^{HR}l/\pi \in [0.35;1.40]$ and $k=2\pi f/c_0$. The positions of the poles are marked for $k^{HR}l/\pi=1.14$ (yellow dots) and the underlying colormap image corresponds to $k^{HR}l/\pi=0.35$. 
(b) Theoretical with losses (dashed curves), experimental (continuous curves) transmission (yellow curves) and reflection (blue curves) coefficients vs frequency for $k^{HR}l/\pi=1.17$ with $f_{max}=645$~Hz.
Inset :  Eigenvalues of the experimental S-matrix for $k^{HR}l/\pi=1.17$. 
(c) Absorption (gray), transmission (orange) and reflection (blue) coefficients at $f_{max}$ vs $k^{HR}l$ (curves for theory and dots for experiments). } 
 \label{fig:serie} 
\end{figure}
We now pay our attention to mirror symmetric scatterers. The interest of these scatterers, compared to the point symmetric scatterers, relies on the fact that the one-sided absorption $\alpha$ takes value larger than $0.5$. This happens since $\alpha_a$ can be different from zero.
Our setup consists of two tuned HRs ($f^{HR}_{1}=f^{HR}_{2}=f^{HR}$) located at different axial positions and
separated by the distance $l$ [see the sketch in Fig. \ref{fig:serie}(a)].  
Now, the detuning parameter is defined as $k^{HR}l=2\pi f^{HR} l /c_0$ where $c_0$ is the sound velocity. 
As in the point symmetric scatterer case, we first inspect the behavior of the lossless transmission coefficient in the complex frequency plane.
In addition to the two HR-related poles, multiple poles due to the Fabry-P\'erot resonances of the waveguide appear. Once the resonance frequency of the HRs is close to a Fabry-P\'erot frequency $f^{FP}$ ($k^{HR}l \sim n \pi$, where $n\in \mathbb{N}$ including $n=0$), one of the poles approaches the real axis and gives rise to an EIT-like mode. In Fig. \ref{fig:serie}(a), the case where $k^{HR}l\rightarrow\pi$ is shown. 
As before, the viscothermal losses can subsequently reduce the amplitude of the transparent peak associated to the EIT-like mode, see Figs. \ref{fig:serie}(b) and (c).
The discrepancies between experimental results and theoretical predictions, larger than in the case of point symmetric scatterers, are attributed to a larger leakage outside the waveguide and to the difficulty to get $f^{HR}_1=f^{HR}_2$ in experiments. We continue our analysis by studying some particular detuning parameters values. For instance, we experimentally (theoretically) find that $r(f_{CPA})\simeq -t(f_{CPA})$ for $k^{HR}l/\pi=0.82$ and $1.17$ ($k^{HR}l/\pi=0.86$ and $1.14$) at $f_{CPA}$. According to the theory, these cases correspond to $\alpha_s(f_{CPA})=1$, which is equivalent to a symmetrical CPA point, $\xi_1(f_{CPA})=0$. 
In Fig. \ref{fig:serie}(b), we verify the existence of a symmetrical CPA point by plotting the scattering coefficients, as well as the eigenvalues of the experimental S-matrix, as a function of frequency for $k^{HR}l/\pi=1.17$. Importantly, at these detuning parameter values, both in theory and experiments, the one-sided absorption $\alpha$ takes a value larger than $0.5$ around $f_{CPA}$. 
Indeed, as we mentioned above $\alpha_a \neq 0$, which is confirmed experimentally by the fact that  $\xi_{2}\neq1$, see inset of Fig. \ref{fig:serie}(b). It is worth to comment here that, by using HRs of smaller Q factor, we observe a larger value of $\alpha$. Overall, we find that the maximum peak of absorption is $\alpha(f_{max})=0.55$ in theory [$\alpha(f_{max})=0.6$ in experiments] for $k^{HR}l/\pi=0.86$ and $1.14$ as shown in Fig. \ref{fig:serie}(c); as before, we show that this corresponds to the critical coupling of the relevant EIT-like mode \cite{supplement}. The maximum peak of absorption appears theoretically at the symmetrical CPA point, i.e., for the same detuning parameter and at  $f_{max}=f_{CPA}$. This is explained by the experimentally observed nearly constant behaviour of  $\xi_{2}$, and thus $\alpha_a$, which is not the case of asymmetric scatterers as we will see below.

\begin{figure}[ht!]
\includegraphics[width=8.6cm]{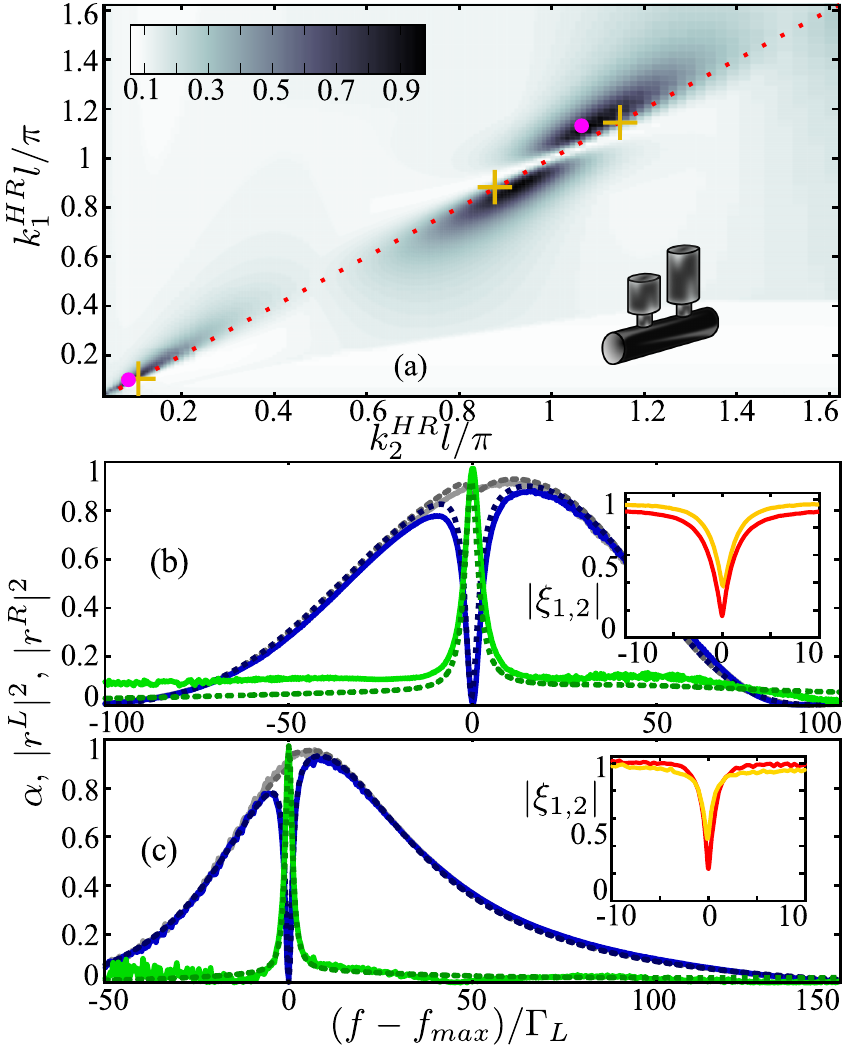}
\caption{(Color online) Asymmetric scaterrer. 
(a) Theoretical absorption vs $k^{HR}_1l/\pi$ and $k^{HR}_2l/\pi$. The red dashed line corresponds to the mirror symmetric scatterers. 
(b)-(c) Theoretical (dashed curves) and experimental (continuous curves) absorption (green), reflection from left (blue) and from right (gray) coefficients. Inset :  Eigenvalues (orange and red curves) of the experimental S-matrix. 
(b) $l=30$~cm, $k^{HR}_1l/\pi=1.14$, $k^{HR}_2l/\pi=1.06$ and $f_{max}=618$~Hz. (c) $l=5$~cm, $k^{HR}_1l/\pi=7.7.10^{-2}$, $k^{HR}_2l/\pi=7.2.10^{-2}$  and $f_{max}=244$~Hz. 
}
 \label{fig:asym} 
\end{figure}
To enhance the one-side incident wave absorption, let us now turn to the case of asymmetric scatterers. We consider an additional degree of freedom in the setup:
 two detuned HRs ($f^{HR}_1\neq f^{HR}_2$) separated by the distance $l$ [see the sketch in Fig. \ref{fig:asym}(a)]. The mirror symmetry is broken ($r^{L}\neq r^{R}$) and two detuning parameters are defined as 
 $k^{HR}_1l$ and $k^{HR}_2l$.  Figure \ref{fig:asym}(a) shows the maximum of one-sided absorption found at $f_{max}$, which is defined with the left incoming wave $\alpha=1-|r^L|^2-|t|^2$,  as a function of the two detuning parameters, where the red dashed line corresponds to the mirror symmetric scatterer case.
Close to the maximum peak of absorption for mirror symmetric scatterers (highlighted by crosses), there exist asymmetric scatterers presenting an Unidirectional Near Perfect One-sided Absorption (UNPOA).
Figures \ref{fig:asym}(b) and (c) show the absorption, the right and left reflection coefficients as a function
of frequency for two cases of UNPOA [highlighted by magenta points in Fig. \ref{fig:asym}(a)]. 
Figure \ref{fig:asym}(b) corresponds to the case where the absorption attains $0.98$ with $k^{HR}_1l/\pi=1.14$ and $k^{HR}_2l/\pi=1.06$. More interestingly, the absorption is $0.96$ for $k^{HR}_1l/\pi=0.08$ and $k^{HR}_2l/\pi=0.07$ in Fig. \ref{fig:asym}(c). This latter case is particularly appealing because it reveals the possibility of UNPOA with a deep sub-wavelength structure (in the experiment, $f_{max}=244$~Hz and $l=5$~cm corresponding to $\lambda/28$). 
Note also that $|r^R| > 0.9$ and $r^L\simeq 0$ near $f_{max}$. This clearly demonstrates the unidirectional character of the absorber.
Besides, it is observed that both eigenvalues of the scattering matrix $\xi_1$ and $\xi_2$ go to near zero values
at $f_{max}$ [see insets in Figs. \ref{fig:asym}(b) and (c)]. 
This differs importantly from the eigenvalues observed with the point symmetric and mirror symmetric scatterers where only one of them is going to zero.

In conclusion, we show that, by the interplay of EIT-like modes and inherent losses, we can obtain strong absorption. In the case of asymmetric configurations, an unidirectional near perfect absorption is possible at frequencies where the incoming wavelength is much larger than the size of the cell. 
Thus, the control of absorption can be achieved by  geometrically tuning the interaction between resonators for a given amount of inherent losses. 
The approach we are following and the resulting phenomenology are not restricted to acoustics, and thus, our findings could find applications to all the domains of wave physics, from matter to electromagnetic and optical waves.

\begin{acknowledgments}
This work has been supported by the Metaudible project ANR-13-BS09-0003, co-funded by ANR and FRAE.
\end{acknowledgments}

\bibliography{2014_Helmholtz_Critical_coupling_bib_v2.7}

\end{document}